\documentclass[preprint,prc,aps,showpacs,showkeys,groupedaddress,floatfix,natbib]{revtex4-1}
\usepackage{epsfig}
\usepackage{dcolumn}
\usepackage{bm}
\usepackage[latin5]{inputenc}
\usepackage{graphics}
\usepackage{graphicx}
\usepackage{epstopdf}
\usepackage{epsfig}
\usepackage{amssymb}
\usepackage{amsmath}
\begin{document}
\title{Properties of an $\alpha$ particle in a Bohrium $270$ Nucleus under the Generalized Symmetric Woods-Saxon Potential}
\author{B.C. L\"{u}tf\"{u}o\u{g}lu}
\affiliation{Department of Physics, Akdeniz University, 07058 Antalya, Turkey}
\email{bclutfuoglu@akdeniz.edu.tr}
\author{M. Erdogan}
\affiliation{Department of Physics, Namik Kemal University, 59030 Tekirdag, Turkey}
\date{\today}
\begin{abstract}

The energy eigenvalues and the wave functions of an $\alpha$ particle in a Bohrium $270$ nucleus were calculated by solving Schrödinger equation for Generalized Symmetric Woods-Saxon potential. Using the energy spectrum by excluding and including the quasi-bound eigenvalues, entropy, internal energy, Helmholtz energy, and specific heat, as functions of reduced temperature were calculated. Stability and emission characteristics are interpreted in terms of the wave and thermodynamic functions. The kinetic energy of a decayed $\alpha$ particle was calculated using the quasi-bound states, which is found close to the experimental value.

\end{abstract}
 \maketitle 
\section{Introduction} \label{sec:intr}

In the last decade, thermodynamic functions have been a subject of ongoing interest in understanding physical properties of potential fields in relativistic or non-relativistic regimes. Pacheco \emph{et al.} has investigated Dirac oscillator in a thermal bath in one-dimension \cite{Pacheco2003}, then extended the study to three-dimensional case \cite{Pacheco2014}. In these studies, for high temperatures, it has been reported that, in the one-dimensional case the heat capacity of the Dirac oscillator is twice as that of one-dimensional harmonic oscillator, while in the three-dimensional case, the limiting value of the specific heat capacity at high temperatures is three times greater than that of the one-dimensional case. Meanwhile, the first experimental one-dimensional Dirac oscillator has been studied by Franco-Villafa\~ne \emph{et al.} \cite{Villafane2013}. Boumali has used the Hurwitz zeta function to investigate the relativistic harmonic oscillator in thermodynamic point of view \cite{Boumali2015}. He has also calculated some thermodynamic functions of graphene under a magnetic field via the two-dimensional Dirac oscillator in an approach based on the zeta function \cite{Boumali2014}. On his following paper, Boumali has studied the thermal properties of the one-dimensional Duffin-Kemmer-Petiau oscillator and computed the vacuum expectation value of its energy by using Hurwitz zeta function \cite{Boumali2015dkp}. Arda \emph{et al.} have studied some thermodynamic quantities of a linear potential in the Klein-Gordon equation with Lorentz vector and Lorentz scalar parts and for an inverse-linear potential, in Dirac equations with a Lorentz scalar term only. In both cases they have given the analytical results for high temperatures under the assumption of strong scalar potential term \cite{Arda2015}. Dong \emph{et al.} have exactly solved a one-dimensional Schrödinger equation of a harmonic oscillator with an additional inverse square potential by using operator algebra. They have studied the relations between the eigenvalues and eigenfunctions using a hidden symmetry and derived some of the thermodynamic functions of the system \cite{Dong2006}.

The Woods-Saxon Potential (WSP) \cite{WoodsSaxon1954} has been widely used in many areas of physics such as nuclear physics \cite{WoodsSaxon1954, BrandanSatchler1997, ZaichenkoOlkhovskii1976, PereyPerey1968, SchwierzWiedenhoverVolya, MichelNazarewiczPloszajczakBennaceur2002, MichelNazarewiczPloszajczak2004, EsbensenDavids2000}, atom-molecule physics \cite{BrandanSatchler1997, Satchler1991}, relativistic \cite{Kennedy2002, PanellaBiondini2010, AydogduArda2012, GuoSheng2005, GuoZheng2002, RojasVillalba2005, HassanabadiMaghsoodi2013, YazarlooMehraban2016, Chargui2016} and  non-relativistic \cite{PahlavaniAlavi2012, CostaPrudente1999, Fluge1994, NiknamRajabi2016} problems.
In order to take the effects such as non-zero $l$, spin-orbit coupling, large force suffered by nucleons near the surface of a nucleus,  additional terms to WSP have been introduced to form various types of Generalized Symmetric Woods-Saxon Potential (GSWSP) \cite{CandemirBayrak2014, BayrakSahin2015, BayrakAciksoz2015, LutfuogluAkdeniz2016, LiendoCastro2016, BerkdemirBerkdemir2005, BadalovAhmado2009,GonulKoksal2007, KouraYamada2000, CapakPetrellis2015, CapakGonul2016, IkotAkpan2012, IkhdairFalayeHamzavi2013}. GSWSP can be used to model any system, in which a particle is trapped in a finite space \cite{surface1, surface2, surface3, surface4, surface5}.

In this paper, we solve Schr\"odinger equation, substituting GSWSP for WSP to calculate the thermodynamic properties of an $\alpha$ particle in a Bohrium $270$ nucleus, as an application of the formalism which has been studied in detail in \cite{LutfuogluAkdeniz2016}. 

Like WSP, the GSWSP does not possess analytical solutions for $l \neq 0$ cases. GSWSP serves our purpose for the case $l=0$, which corresponds to spherical symmetry. This reduces the problem to one-dimensional form with the only radial degree of freedom.

In section \ref{sec:materyel}, we first consider the GSWSP, and give the main result of \cite{LutfuogluAkdeniz2016}, in section \ref{sec:bulgu} we calculate the energy spectrum of an $\alpha$ particle employing the method given in the previous section. In subsection \ref{thermosection}, we calculate partition functions using the energy spectra, then, Helmholtz free energies, internal energies, entropies and specific heat capacities of the system as functions of temperature. In section \ref{sec:d}, our conclusion is presented.

\section{Material and Method} \label{sec:materyel}

Let us consider a nucleon under one-dimensional GSWSP \cite{LutfuogluAkdeniz2016};
\begin{eqnarray}\label{gws}
  V(x)&=&\theta{(-x)}\Bigg[-\frac{V_0}{1+e^{-a(x+L)}}+\frac{W_0 e^{-a(x+L)}}{\big(1+e^{-a(x+L)}\big)^2}\Bigg] \nonumber \\
  &+& \theta{(x)}\Bigg[-\frac{V_0}{1+e^{a(x-L)}}+\frac{W_0 e^{a(x-L)}}{\big(1+e^{a(x-L)}\big)^2}\Bigg], \label{gsws}
 \end{eqnarray}
here the second terms in the square brackets represent the energy barrier at the surface, which is linearly proportional to the spatial derivative of the first term and the radius. Thus the parameter $W_0$ is linearly proportional to $a, L, V_0$ and the proportionality constant can be determined by means of momentum and energy conservations for the nucleus under consideration. $\theta{(\pm x)}$ are the Heaviside step functions, $V_0$ is the depth of the potential given by \cite{PereyPerey1968}
\begin{eqnarray}
  V_0 &=& 40.5+ 0.13 A.
\end{eqnarray}
We classify bound states to tight-bound and quasi-bound states since they obey different boundary conditions. In tight-bound case, the particles are confined in the well and they can have only negative energy eigenvalues. In other words their wave functions outside the well vanish.  In quasi-bound case, although the particles are inside the potential well, they have positive energy eigenvalues and with appropriate conditions they can tunnel.  Therefore their wave functions imply a propagation in outgoing direction from the potential well, contrary to tight-bound states. Exploiting the continuity of the wave functions and their first spatial derivatives, as well behaved wave functions must obey, the quantization conditions are obtained. Moreover by using the $x\rightarrow -x$ symmetry of the potential well the energy spectrum can be given in two subsets as "even" and "odd" eigenvalues. In the reference \cite{LutfuogluAkdeniz2016} the energy eigenvalues for tight-bound states have been found as
\begin{eqnarray}
E_n^{tb-e}&=&-V_0+\frac{\hbar^2}{2 m L^2}\Bigg|\arctan \frac{(N_1-N_2)}{i(N_1+N_2)}\pm n'\pi\Bigg|^2, \,\,\,\,\,\,\, \label{tbeven}\\
E_n^{tb-o}&=&-V_0+\frac{\hbar^2}{2 m L^2}\Bigg|\arctan \frac{(N_1+N_2)}{i(N_1-N_2)}\pm n'\pi\Bigg|^2,\label{tbodd}
\end{eqnarray}
while for quasi-bound states are
\begin{eqnarray}
E_n^{qb-e}&=&-V_0+\frac{\hbar^2}{2 m L^2}\Bigg|\arctan \frac{(N_3-N_4)}{i(N_3+N_4)}\pm n'\pi\Bigg|^2, \,\,\,\,\,\,\, \label{qbeven}\\
E_n^{qb-o}&=&-V_0+\frac{\hbar^2}{2 m L^2}\Bigg|\arctan \frac{(N_3+N_4)}{i(N_3-N_4)}\pm n'\pi\Bigg|^2.\label{qbodd}
\end{eqnarray}
Here $n'$ are integers, whereas $n$ stands for the number of nodes, the roots of the wave functions. $N_1$, $N_2$, $N_3$, and $N_4$ are complex numbers
\begin{eqnarray}
  N_1 &=& \frac{\Gamma(c_1)\Gamma(c_1-a_1-b_1)}{\Gamma(c_1-a_1)\Gamma(c_1-b_1)}, \\
  N_2 &=& \frac{\Gamma(c_1)\Gamma(a_1+b_1-c_1)}{\Gamma(a_1)\Gamma(b_1)}, \\
  N_3 &=& \frac{\Gamma(2-c_1)\Gamma(c_1-a_1-b_1)}{\Gamma(1-a_1)\Gamma(1-b_1)}, \\
  N_4 &=& \frac{\Gamma(2-c_1)\Gamma(a_1+b_1-c_1)}{\Gamma(1+a_1-c_1)\Gamma(1+b_1-c_1)},
\end{eqnarray}
and implicitly dependent on the energy eigenvalues via the coefficients $a_1$, $b_1$ and $c_1$
\begin{eqnarray}
  a_1 &=& \mu+\theta+\nu, \\
  b_1 &=& 1+\mu-\theta+\nu, \\
  c_1 &=& 1+2\mu,
\end{eqnarray}
where
\begin{eqnarray}
  \mu       &=& \sqrt{-\frac{2 m E_n}{a^2 \hbar^2}}, \\
  \nu       &=& \sqrt{-\frac{2 m (E_n+V_0)}{a^2 \hbar^2}}, \\
  \theta    &=& \frac{1}{2}\mp\sqrt{\frac{1}{4}-\frac{2mW_0}{a^2\hbar^2}}.
\end{eqnarray}
\section{Results}\label{sec:bulgu}

\begin{figure}[ht]
\centering
\includegraphics[width=.7\linewidth]{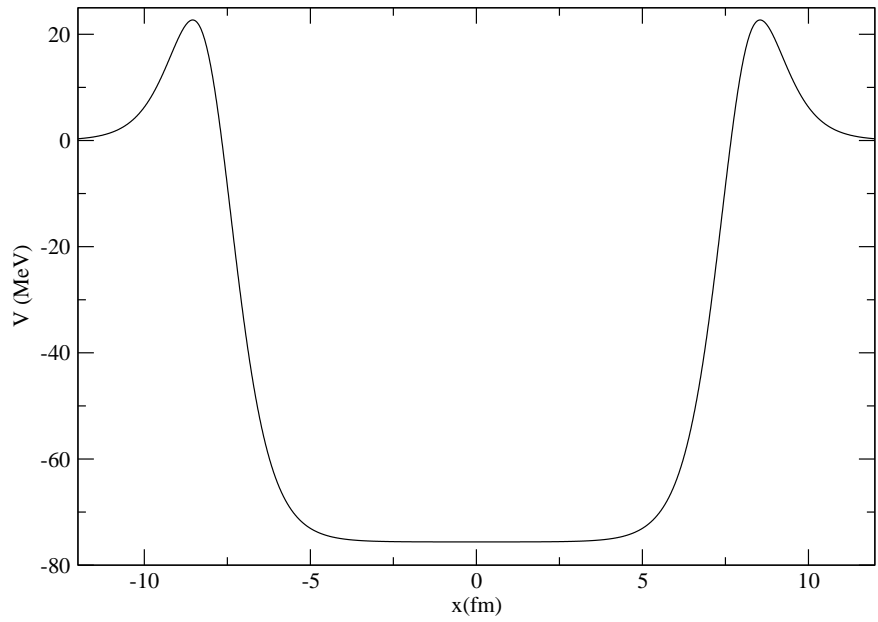}
\caption{The GSWS potential for an $\alpha$-particle in a Bohrium $270$ nucleus.}
\label{fig:Figure_1}
\end{figure}

In this manuscript we investigate an $\alpha$ particle in a heavy nucleus, Bohrium $270$. $a$, the reciprocal of the diffusion parameter, is taken to be $a=1.538 fm^{-1}$ \cite{PereyPerey1968} and the nuclear radius is calculated as $L=8.068 fm$.  Then we substitute the atomic number $A=270$ of the nucleus, and find out $V_0=75.617 MeV$ and $W_0=215.523 MeV$. In Figure \ref{fig:Figure_1}, the GSWSP is shown. 

The tight-bound energy eigenvalues calculated via Equation \ref{tbeven} and Equation \ref{tbodd}  fall into the interval such that, 
\begin{eqnarray}
  -V_0 <E_n^{tb}<0,
\end{eqnarray}
are presented in Table \ref{energy_spectrum_tb}.
\begin{table}[h]
\centering
\caption{The tight-bound energy spectrum of the $\alpha$ particle confined in Bohrium $270$ nucleus.} \label{energy_spectrum_tb}
\begin{tabular}{|c|c|c|c|c|c|}
\hline
$n$ &$E_n^{tb}(MeV)$ & $n$ &$E_n^{tb}(MeV)$& $n$ &$E_n^{tb}(MeV)$\\
\hline
$0$&$-75.166$&$6$&$-59.531$&$12$&$-29.605$\\
\hline
$1$&$-73.915$&$7$&$-55.373$&$13$&$-23.607$\\
\hline
$2$&$-72.022$&$8$&$-50.855$&$14$&$-17.386$\\
\hline
$3$&$-69.585$&$9$&$-45.998$&$15$&$-10.971$\\
\hline
$4$&$-66.666$&$10$&$-40.823$&$16$&$-4.402$\\
\hline
$5$&$-63.304$&$11$&$-35.351$&$$&$$\\
\hline
\end{tabular}
\end{table}
The quasi-bound energy eigenvalues given by Equation \ref{qbeven} and Equation \ref{qbodd} satisfy
\begin{eqnarray}
  0<E_n^{qb}<V_0\frac{(1-aL)^2}{4a}
\end{eqnarray}
i.e. $0<E_n^{qb}<22.705 MeV$, which are given in Table \ref{energy_spectrum_qb}.
\begin{table}[h]
\centering
\caption{The quasi-bound energy spectrum of the $\alpha$ particle confined in Bohrium $270$ nucleus.} \label{energy_spectrum_qb}
\begin{tabular}{|c|c|}
\hline
$n$ &$E_n^{qb}(MeV)$\\
\hline
$17$&$2.263- 0.537 \times 10^{-3} i$\\
\hline
$18$&$8.929 - 0.146 \times 10^{-1} i$\\
\hline
$19$&$15.439 - 0.133 i$\\
\hline
$20$&$21.688 - 0.650 i$\\
\hline
\end{tabular}
\end{table}
Note that the tight-bound states are considered stationary, since their energy eigenvalues are real with infinite time constants. Contrarily, energy eigenvalues of quasi-bound states have complex form in general, giving rise to a finite time constant and a non-zero decay probability \cite{Gamow1928, Siegert1939}.

\begin{figure}[ht]
\centering
\includegraphics[width=.7\linewidth]{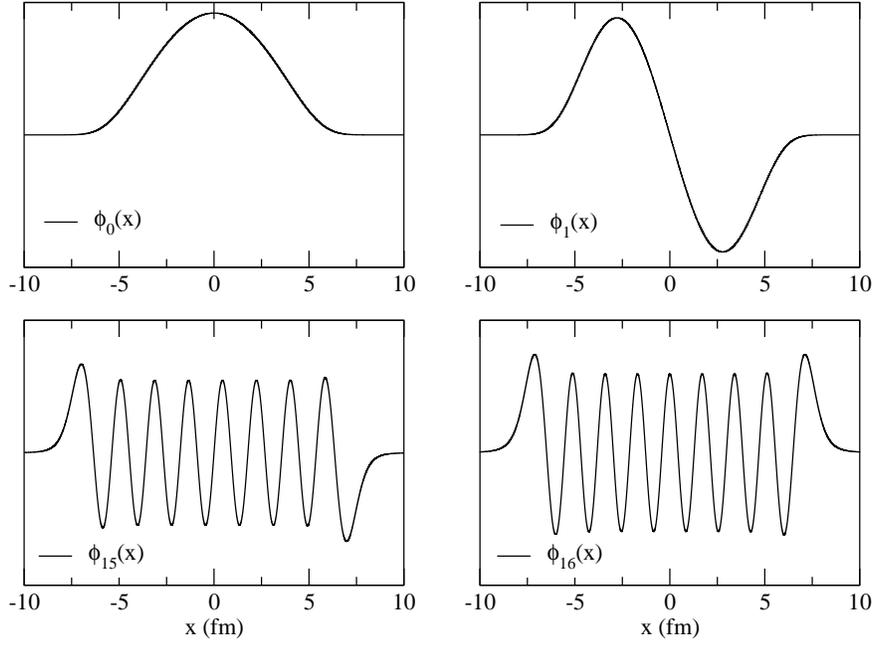}
\caption{The first four unnormalized wave functions of the tight-bound states spectrum.}
\label{fig:Figure_2}
\end{figure}

\begin{figure}[ht]
\centering
\includegraphics[width=.7\linewidth]{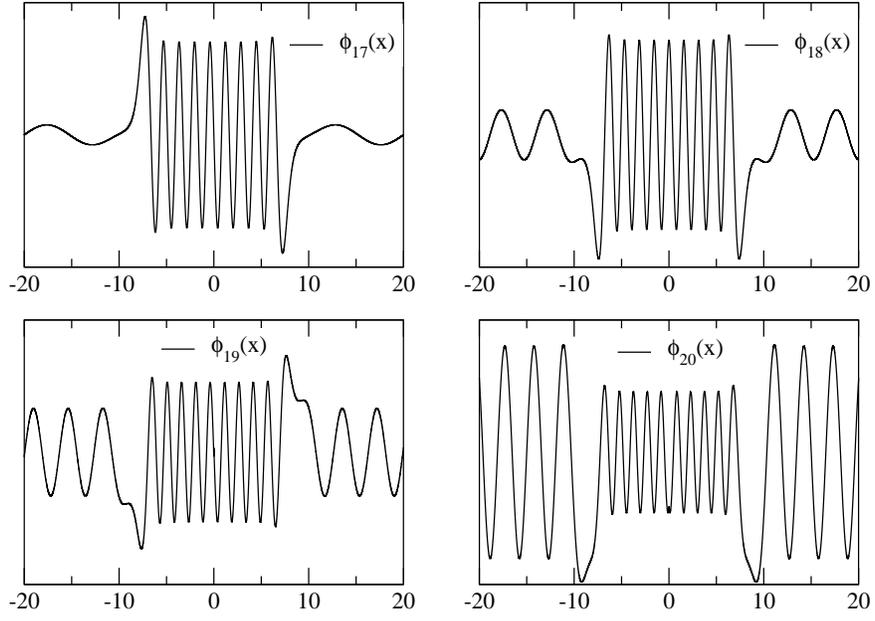}
\caption{The unnormalized wave functions of the quasi-bound states and the oscillations outside the well.}
\label{fig:Figure_3}
\end{figure}

The first two and the last two bound state wave functions are given in Figure \ref{fig:Figure_2}, and all the quasi-bound wave functions are shown in Figure \ref{fig:Figure_3}. The wave functions of the quasi-bound states have oscillations outside the well, indicating the $\alpha$ decay. Note that all the wave functions are unnormalized.

\subsection{Thermodynamic functions of the system}\label{thermosection}

\begin{figure}[ht]
\centering
\includegraphics[width=.7\linewidth]{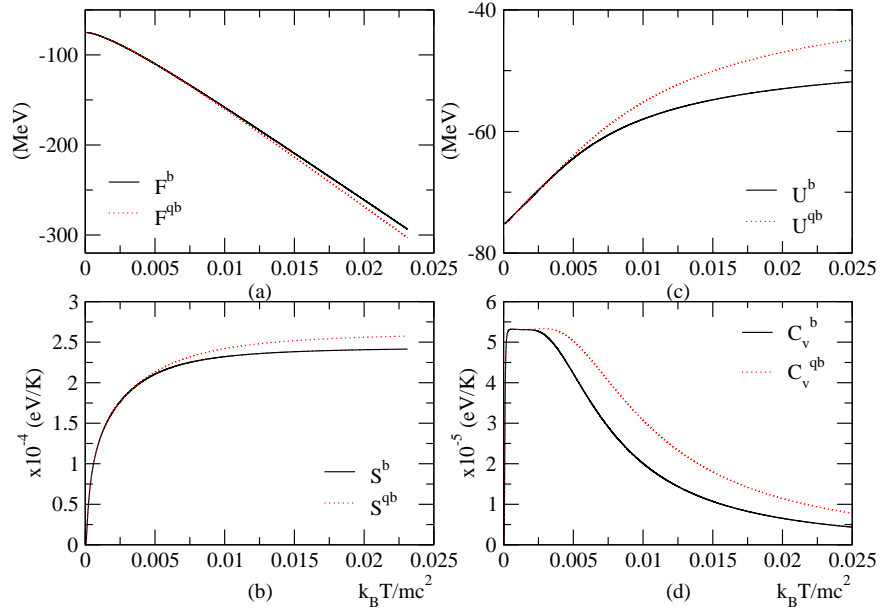}
\caption{(a) Helmholtz free energy, (b) entropy, (c) internal energy, (d) specific heat capacity versus reduced temperature $k_BT/mc^2$. Quasi bound states are included in the red curve, excluded in the black. The saturation value of the internal energy is $-46.739 MeV$ without quasi bound states, and $-35.535 MeV$, when they are included.}
\label{fig:Figure_5}
\end{figure}

We first calculate the partition function by using the energy spectrum of the system;
\begin{eqnarray}
  Z(\beta) &=& \sum_{n=0} e^{-\beta E_n},
\end{eqnarray}
Here $\beta$ is defined by
\begin{eqnarray}
  \beta &=& \frac{1}{k_B T}.
\end{eqnarray}
where $k_B$ indicates the Boltzman constant and $T$ is the temperature in Kelvin. The Helmholtz function of the system is calculated via the relation, 
\begin{eqnarray}
      F(T) &\equiv& -k_B T\ln Z(\beta).
    \end{eqnarray}
The entropy of the system is calculated using,
  \begin{eqnarray}
      S(T) &=& -\frac{\partial}{\partial T}F(T).
    \end{eqnarray}
The Helmholtz free energy and the entropy functions for both quasi-bound states included and excluded of the system versus the reduced temperature are seen in Figure \ref{fig:Figure_5}(a) and Figure \ref{fig:Figure_5}(b), respectively. The reduced temperature is defined as the unitless quantity $k_BT/mc^2$. The zero entropy at zero Kelvin is consistent with the third law of thermodynamics. The entropy saturates to the value $2.33 \times 10^{-4} eV/K$ when only the bound states are included in the partition function. When the quasi-bound states are taken into account, the number of microstates available to the system increases and the entropy saturates to a higher value, $2.96 \times 10^{-4} eV/K$ and the Helmholtz free energy decreases for high temperatures as seen in Figure \ref{fig:Figure_5}(a). According to the behavior of entropy function, the system favours the addition of quasi-bound states, being consistent with the second law of thermodynamics. This verifies the necessity of the surface interactions. 

The internal energy $U(T)$ is the expectation value of the energy of the $\alpha$ particle. It is given by 
\begin{eqnarray}
      U(T) &=& -\frac{\partial}{\partial \beta}\ln Z(\beta).
    \end{eqnarray}
Then, the specific heat capacity $C_v(T)$  is 
\begin{eqnarray}
      C_v(T) &\equiv& \frac{\partial}{\partial T}U(T).
    \end{eqnarray}
For the cases that the quasi-bound states are included and excluded, the plots of $U(T)$ and $C_v(T)$ are shown in Figure \ref{fig:Figure_5}(c) and Figure \ref{fig:Figure_5}(d), respectively. 

\begin{figure}[h]
\centering
\includegraphics[width=.7\linewidth]{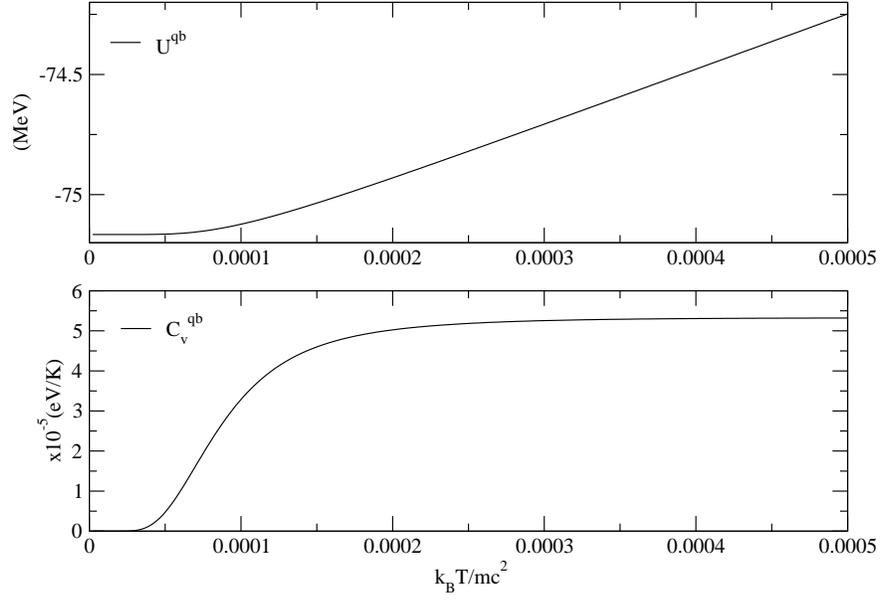}
\caption{The internal energy initially increases in a convex manner followed by a linear ascent.}
\label{fig:Figure_6}
\end{figure}

The initial behavior of the internal energy and the specific heat against the reduced temperature are presented in Figure \ref{fig:Figure_6}. The initial value of the internal energy at $0K$ is $-74.995 MeV$, the lowest energy eigenvalue in the spectrum. The internal energy has an initial convex increase until the reduced temperature of $2\times 10^{-4}$ followed by a linear ascent. The linear increase of the internal energy is followed by a concave ascent up to the saturation, which is $-46.739 MeV$ when the quasi-bound states are excluded, and $-35.535 MeV$ when they are included. These saturations are the mean values of the respective energy spectra. This is a consequence that, as temperature goes to infinity, all Boltzmann factors approach to unity. In this temperature regime, thus occupation of all energy values become equally probable in the spectrum. \\

In the linear regime, the specific heat remains constant at about $5.35 \times 10^{-5} eV/K$. When the quasi-bound are presented in the spectrum, the specific heat goes to zero in a wider temperature range.

\begin{figure}[ht]
\centering
\includegraphics[width=.7\linewidth]{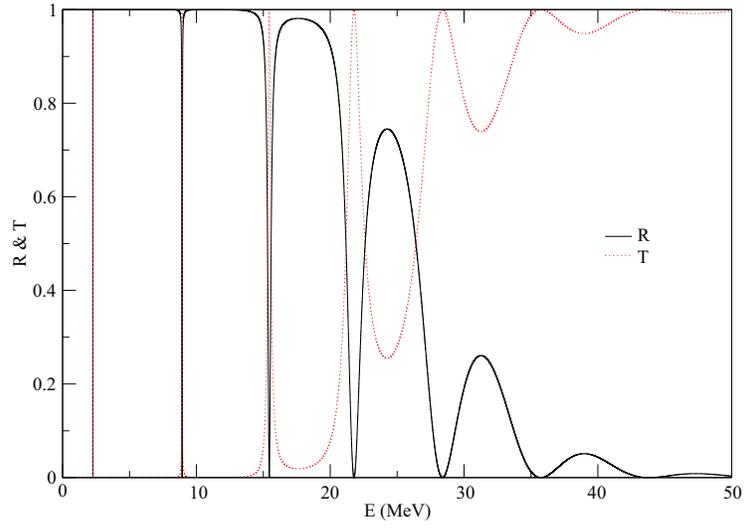}
\caption{Transition ($T$) and reflection ($R$) coefficients of an $\alpha$ particle through the energy barrier at the surface, as functions of the energy.}
\label{fig:Figure_4}
\end{figure}

A particle having the energy that coincides the quasi-bound energy spectrum after the tunneling into a nucleus is known as the resonance. In our problem the resonance is satisfied for the $\alpha$ particle to decay, when it has a quasi-bound energy as indicated in Figure \ref{fig:Figure_4}. The kinetic energy of the decayed $\alpha$ particle has been calculated as the mean value of the quasi-bound energy spectrum as $12.079 MeV$, which is in reasonable agreement with the experimental data \cite{DunfordBurrows1995}.

\section{Discussion and Conclusion} \label{sec:d}

In this work, we revealed the contribution of the surface term added to WSP to form GSWSP, to the physics of a Bohrium $270$ nucleus, in context of thermodynamic point of view. We solved the Schr\"odinger equation considering an $\alpha$ particle in a GSWSP, which has surface terms in addition to the WSP. The energy spectrum and the corresponding wave functions of the system were calculated as well as entropy, internal energy, Helmholtz energy, and specific heat, as functions of reduced temperature, using the partition functions based on the energy spectrum.\\

When the quasi-bound states are taken into account, the internal energy increases, while the Helmholtz energy decreases in comparison with the case of bound states solely. With this inclusion, the entropy also increases and the specific heat capacity sails at higher values, decaying to zero at longer temperature scale. The bound state wave functions imply that the nucleon is completely restricted within the nucleus, with zero decay probability, while decay probabilities have resonances leading to very high tunneling probabilities for quasi-bound states. The imaginary parts of the quasi-bound energy eigenvalues are used to calculate the kinetic energy of the decayed $\alpha$ particle, being in reasonable agreement with the experimental data. The difference from the experimental data is because of that the other effects such as spin-orbit coupling, orbital contribution were ignored, which reduces the original problem into one-dimensional one.

\section*{Acknowledgment} \label{sec:e}

This work was partially supported by the Turkish Science and Research Council (T\"{U}B\.{I}TAK) and Akdeniz University. The authors would like to thank to Dr. E. Pehlivan for his valuable assistance and to the unknown correspondence for scientific discussions throughout this work.

\newpage

\newpage

\newpage

\newpage

\newpage

\newpage

\end{document}